# Non-magnetic impurity scattering in a $d_{x^2-y^2}$ superconductor near a van Hove point: Zn versus Ni in the cuprates

R. Fehrenbacher

*Max-Planck-Institut für Festkörperforschung, Heisenbergstr. 1, D-70569 Stuttgart, Germany*

(December 15, 1995)

We consider the effect of non-magnetic impurities in a $d_{x^2-y^2}$ superconductor with $\varepsilon_F$ close to a van Hove singularity. It is shown that the non-trivial density of states (DOS) allows for resonant scattering already at intermediate potential strengths $|u| \approx 1 - 2$eV. The residual DOS at $\varepsilon_F$, and the $T_c$ suppression rate are found to strongly depend on the carrier concentration. Quantitative agreement with experiments on Zn and Ni doped cuprates is obtained by adjusting a single parameter, $u$.

The effect of non-magnetic impurities on the properties of a superconductor (SC) can provide useful information about the symmetry of its order parameter (OP). Anderson's theorem [1] states that non-magnetic impurities neither affect the transition temperature $T_c$, nor the density of states (DOS) $N(\omega)$ of a BCS SC with an *isotropic* OP. However, if the OP is anisotropic, possibly exhibiting nodes (such as e.g. a $d_{x^2-y^2}$ OP), non-magnetic impurities may lead to a large residual DOS $N_{res} \equiv N(\omega = 0)$ at the Fermi energy $\varepsilon_F$, and even to a complete suppression of $T_c$ [2].

In the high-$T_c$ copper-oxides, Zn impurities are particularly harmful. For YBa$_2$(Cu$_{1-x}$Zn$_x$)$_3$O$_{7-y}$ for instance, where Zn substitutes primarily on *planar* Cu sites [3], a rapid suppression of $T_c$ ($T_c = 0$ at $x \approx 0.08 - 0.1$), as well as a $N_{res}$ growing like $x^{1/2}$ is observed [4]. On the other hand, nominally magnetic Ni$^{+2}$ impurities (also believed to primarily substitute for planar Cu) have a much milder effect: while suppressing $T_c$ at a rate 1/2 - 1/3 of Zn, they do not lead to a significant $N_{res}$ up to $x \approx 0.05$ [4].

An appealing and simple interpretation of these results is possible in terms of a $d_{x^2-y^2}$ OP affected by non-magnetic scattering, assuming that Zn acts as a strong, in fact resonant scatterer [5]. However, one problem with this model comes from the fact that the growth in the normal state resistivity $d\rho_n/dx$ just above $T_c$ is quite similar for Ni and Zn [6]. Given that, their different behavior in the SC state is puzzling, since in standard models (for magnetic *and* non-magnetic impurities) [7,8], the $T_c$ suppression rate $dT_c/dx$, and the residual resistivity are both determined by the same parameter, the normal state scattering rate. Also, it is not clear why Zn should scatter resonantly, since this is obtained only for scattering strength $|u| \to \infty$.

Reports about localized magnetic moments on Cu sites next to a Zn [9] further complicate the issue. The size of the moment depends on hole doping, and seems to be large in underdoped cuprates ($\approx 0.9 - 1.4\mu_B$) [9, 10], whereas in optimally doped YBa$_2$Cu$_3$O$_{7-y}$ (Y-O$_7$) it is small [9] or maybe absent [11]. Therefore, at least in Y-O$_7$, it seems justified to model Zn as a pure potential scatterer. In underdoped samples, additional magnetic pair breaking might occur. Note however, that even then it is not obvious that Abrikosov-Gorkov (AG) theory [7] applies, due to (i) the existence of short-range antiferromagnetic (AF) correlations, and (ii), the fact that the 'impurity' spins are probably not totally immobile, but part of the same SC spin fluid (Cu spin 1/2). For a strongly correlated system like the *t-J* model for instance, it was shown [12], that a localized spin 1/2 impurity only has a small influence on magnetic and pairing properties when its coupling to the mobile spins $J_i$ is the same or similar to the coupling $J$ among the latter. This should be satisfied here. There is also the possibility that Zn might act on the *pair potential* itself as proposed recently [13] in the context of an AF spin fluctuation model. However, the presence of localized Cu moments nearest neighbour to Zn suggests that AF correlations are probably unaffected or maybe even enhanced in the vicinity of the impurity site, since this region corresponds to a locally underdoped phase, *i.e.*, is closer to AF.

As one possible route to resolve the discrepancies between theory and experiment described above, we explore the influence of a realistic quasiparticle dispersion on the effect of non-magnetic impurity scattering in a model BCS superconductor with an assumed $d_{x^2-y^2}$ OP chosen as $\Delta_{\mathbf{k}} = \Delta\eta_{\mathbf{k}}$, with $\eta_{\mathbf{k}} = (\cos k_x - \cos k_y)/2$. Our main findings are: (i) the existence of a van Hove singularity (vHs) just below $\varepsilon_F$ (suggested by angular-resolved photoemission [ARPES] [14, 15]) leads to a strong violation of particle-hole symmetry, and allows for the occurrence of resonant scattering, at realistic potential strengths $|u| \approx 1 - 2$eV. (ii) The *sign* of $u$ is crucial. For hole concentrations corresponding to Y-O$_7$, we find that resonant scattering occurs only for negative $u$ (in electron notation). A large negative $u$ is expected for Zn impurities, since an inert $d$-shell strongly repels holes, *i.e.*, attracts electrons. This is also consistent with the observed localized Cu moments. Ni, however, should create a weaker *attractive* potential for holes, *i.e.*, positive $u$, since a Ni$^{+III}$ oxidation state is

1

formed more easily than Cu$^{+III}$. (iii) The vHs also results in a *strong frequency* dependence of the normal state quasiparticle lifetime. This might explain why $T_c$ suppression rate and increase of residual resistivity are not simply related as observed for Zn and Ni. (iv) Both, $N_{res}$, and $dT_c/dx$ are strongly doping dependent, the latter becoming much larger in the underdoped regime away from the vHs. (v) Overall, good *quantitative* agreement with experiments is obtained with only one adjustable parameter, $u$.

We model the impurities by a short-range potential $V(\mathbf{r}) = u\delta(\mathbf{r} - \mathbf{r}_i)$, and apply the self-consistent t-matrix approximation [8, 16]. This approach has recently been shown to yield accurate results in the dilute impurity limit [17]. Working in particle-hole space, the t-matrix $\widehat{T}$ satisfies the following Lippmann-Schwinger equation (quantities with a hat represent matrices)

$$\widehat{T}(\omega) = u\left[\widehat{\sigma}_3 + \widehat{\sigma}_3\widehat{T}(\omega)\frac{1}{\Lambda}\sum_{\mathbf{k}}\widehat{g}(\mathbf{k},\omega)\right]. \quad (1)$$

Here we introduced the propagator

$$\widehat{g}(\mathbf{k},\omega) = \frac{\widetilde{\omega}\widehat{\sigma}_0 + \widetilde{\Delta}_{\mathbf{k}}\widehat{\sigma}_1 + \widetilde{\xi}_{\mathbf{k}}\widehat{\sigma}_3}{\widetilde{\omega}^2 - \widetilde{\Delta}_{\mathbf{k}}^2 - \widetilde{\xi}_{\mathbf{k}}^2}, \quad (2)$$

where $\widetilde{\omega} = \omega - \Sigma_0, \widetilde{\Delta}_{\mathbf{k}} = \Delta_{\mathbf{k}} + \Sigma_1, \widetilde{\xi}_{\mathbf{k}} = \xi_{\mathbf{k}} + \Sigma_3$, $\xi_{\mathbf{k}}$ the quasiparticle energy. The self-energy (like all other matrices) is expanded in terms of the identity and Pauli matrices $\widehat{\sigma}_0 \cdots \widehat{\sigma}_3$ as $\widehat{\Sigma} = \Sigma_j \widehat{\sigma}_j$, and is given by $\widehat{\Sigma} = n_i\widehat{T}$, $n_i$ the impurity concentration, and $\Lambda$ the volume. For the $d_{x^2-y^2}$ OP, the off-diagonal self-energy vanishes by symmetry, and from (1), the non-zero components are

$$\Sigma_0 = \frac{n_i G_0}{(1/u - G_3)^2 - G_0^2}; \quad \Sigma_3 = \frac{n_i(1/u - G_3)}{(1/u - G_3)^2 - G_0^2}. \quad (3)$$

where $G_i(\omega) = \Lambda^{-1}\sum_{\mathbf{k}} g_i(\mathbf{k},\omega)$. Previous approaches [5, 8] assumed particle-hole symmetry in the normal DOS, which has the consequence that $\Sigma_3 = 0$ is a self-consistent solution of (3). Here, we relax this condition, and take into account the full structure of the t-matrix. Unfortunately, this also means that all **k**-sums have to be done numerically when iterating Eqs. (3) for a self-consistent solution. A $d \times d$ **k**-space grid with $d = 4000$ for calculations of spectral quantities, and $d = 200$ to calculate $T_c$ proved to be sufficient to obtain good convergence.

For the quasiparticle dispersion $\xi_{\mathbf{k}}$, we use a tight-binding fit to ARPES data on Bi-2212 with real space hopping matrix elements $[t_0, \cdots, t_5] = [0.879, -1, 0.28, -0.087, 0.094, 0.087]$, ($t_0$ on-site, $t_1$ nn, $t_2$ nnn hopping, ...) [18, 19]. All energies are measured in units of $|t_1| = 0.149$eV. The value of $t_0$ corresponds to a hole doping of $\delta = 0.17$. In the calculations, we consider doping levels of $\delta = 0.25, 0.18, 0.13$ ($t_0 = 1.0, 0.9, 0.8$) which are representative for Y-O$_7$ ($\varepsilon_{vHs} \approx -16$meV [15]), Bi-2212 ($\varepsilon_{vHs} \approx -31$meV [14]), and Y-O$_{6.6}$ (no precise value of $\varepsilon_{vHs}$ is published to date) respectively, $\varepsilon_{vHs}$ the position of the vHs with respect to $\varepsilon_F$. A rigid band picture is assumed to be an acceptable approximation in this small doping interval. In the calculations of spectral quantities, we set $\Delta = 0.2$ (30 meV). The *only* free parameter in the theory is then the potential strength $u$, since the impurity concentration $n_i$ is also fixed by experiment. We shall search for values of $u$ to consistently explain the experiments for Zn and Ni.

The effect of the non-trivial DOS is rather dramatic, in particular if a vHs is close to $\varepsilon_F$, as in the present case: (i) The s-wave scattering phase shift $\delta_0$ acquires a strong frequency dependence already in the *normal* state, which also reflects itself in the scattering rate, hence resistivity. Furthermore, the dependence on $u$ is highly non-trivial. (ii) In the self-energy Eqs. (3), the cotangent $c = \cot\delta_0$ (being *frequency independent* in case of a constant normal DOS) that is usually used to parameterize the scattering strength, is replaced by the (now *frequency dependent*) quantity $\widetilde{c}(\omega) \equiv -1/u + G_3(\omega)$. This leads to a strong sensitivity of the superconducting DOS and $T_c$ on both $u$ and the chemical potential. Resonant scattering is usually observed for $|c| \ll 1$ (corresponding to $|u| \approx \infty$ for constant DOS). In the present case, this translates to the condition $\widetilde{c}(\omega = 0) \ll 1$, allowing for resonant scattering even if $|u| \ll \infty$. We shall illustrate these points in the following.

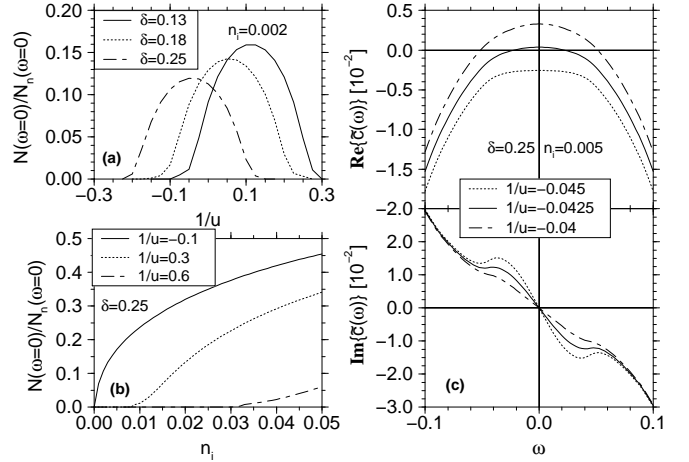

**Fig. 1.** (a) The residual DOS $N_{res}$ as a function of $1/u$ and different doping levels. (b) $N_{res}$ as a function of impurity concentration $n_i$ for various values of $1/u$. (c) The real and imaginary part of $\widetilde{c}(\omega)$ for values of $1/u$ near resonance.

Fig. 1(a) shows $N_{res} \equiv N(\omega = 0) = -\pi^{-1}\text{Im}\{G_0(\omega) + G_3(\omega)\}|_{\omega=0}$, as a function of the potential strength $u$ for fixed $n_i = 0.002$. A broad peak is observed, the center of which shifts from negative ($1/u \approx -0.05$ for $\delta = 0.25$) to positive ($1/u \approx 0.1$ for $\delta = 0.13$) values as the hole doping is lowered. The broadness of the peaks shows that the signature of resonant scattering, *i.e.*, a large $N_{res}$ for very small $n_i$, occurs in a rather extended interval on the $1/u$ axis. Using $\delta = 0.25$ to model Y-O$_7$, we find that a value as small as $1/u = -0.1$, i.e., $|u| \approx 1.5$eV $\approx W$, the bandwidth, is sufficient to explain the values reported for Zn [4]. This assignment immediately leads to the prediction that the Zn induced $N_{res}$ should decrease significantly for underdoped samples, e.g., for $\delta = 0.13, N_{res}(n_i = 0.002) \approx 0$. In Fig. 1(b), we plot the $n_i$ dependence of $N_{res}$ for selected values of $1/u$, and $\delta = 0.25$. The expected relation $N_{res} \sim n_i^{1/2}$ holds for $1/u = -0.1$, whereas for $1/u > 0.6$, $N_{res}$ is negligibly small up to concentrations of $n_i = 0.05$. The latter behavior was reported for Ni doped Y-$O_7$, hence, we assign a value of $1/u \approx 0.7$ to this case. Note, that if substitution occurs primarily on the planar Cu site, $n_i \approx 1.5x$ in $YBa_2(Cu_{1-x}T_x)_3O_{7-y}$ (T = Zn,Ni). However, at larger $x$, substitution is also expected on the Cu(1) chain sites. Fig. 1(c) confirms the point made earlier, that the occurrence of resonant scattering is now linked to the condition $|\tilde{c}(\omega = 0)| \ll 1$. For $\delta = 0.25$, $|\tilde{c}(\omega = 0)| = 0$, i.e., resonance, occurs for $1/u \approx -0.0425$.

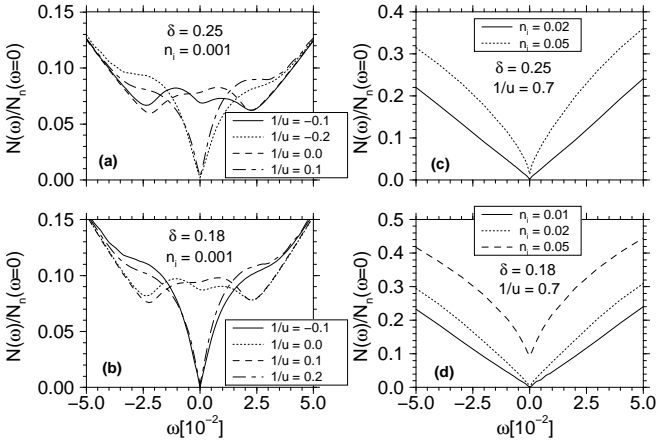

**Fig. 2.** The $\omega$ dependence of the DOS for small $n_i = 0.001$ near resonance (a,b), and for larger $n_i$ far from resonance (c,d).

Figs. 2(a-d) show the frequency dependence of the DOS near $\varepsilon_F$. For very small $n_i$ [Figs. 2(a,b)], we obtain the characteristic hump at $\varepsilon_F$, provided the resonance condition $|\tilde{c}(\omega = 0)| \ll 1$ is satisfied. On the other hand, for weaker scattering $1/u = 0.7$ [Figs. 2(c,d)], $N(\omega)$ vanishes linearly at $\varepsilon_F$ for $n_i \leq 0.02$, and for larger $n_i$, a sublinear dependence is found with a small $N_{res}$, which grows as $\delta$ is reduced. From this, we predict that a small $N_{res}$ could appear in *underdoped* Ni substituted cuprates for $n_i < 0.05$.

To calculate $T_c$, we solve the linearized gap equation in the presence of impurities

$$1 = \frac{V}{4\Lambda\beta_c} \sum_{n,\mathbf{k}} \frac{(\cos k_x - \cos k_y)^2}{\tilde{\omega}_{nc}^2 + \tilde{\xi}_\mathbf{k}^2}. \quad (4)$$

The pair potential is chosen as $V_{\mathbf{kk'}} = V\eta_\mathbf{k}\eta_{\mathbf{k'}}$ to generate a $d_{x^2-y^2}$ OP and $\beta_c = (k_B T_c)^{-1}$. Due to the non-trivial $\xi_\mathbf{k}$, the normal state ($\Delta = 0$) self-energy entering (4) becomes frequency dependent. Hence, $T_c$ cannot be expressed in the standard AG form, and the Matsubara sum has to be evaluated numerically with a cut-off, typically $\omega_n \leq 50$ ($\approx$ 7.5eV). For the pure case, this results in an accuracy better than $10^{-4}$ compared to the value without cut-off. The coupling constant was chosen as $V = 2.2$, which corresponds to a near neighbour attraction of $V/8 = 0.275 \approx$ 41meV (see [18]). Without impurities, this leads to $T_c = (111K, 92K, 65K)$ for $\delta = (0.25, 0.18, 0.13)$.

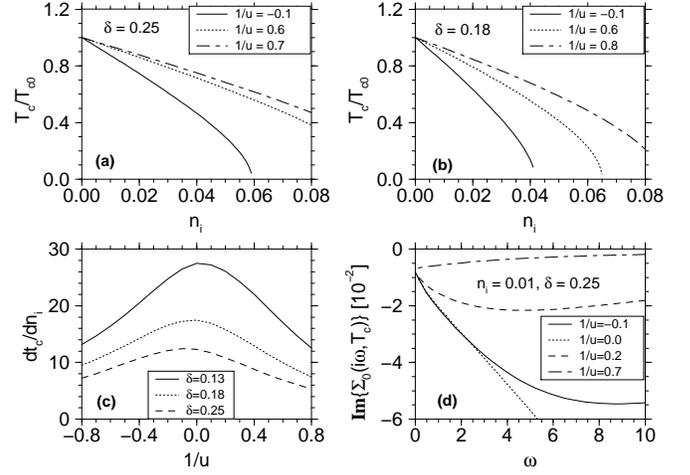

**Fig. 3.** (a,b) The $T_c$ suppression for potentials near and far from resonance, for $\delta = 0.18, 0.25$. (c) $1/u$ dependence of the $T_c$ suppression rate for $\delta = 0.13, 0.18, 0.25$. (d) Frequency dependence of Im$\Sigma_0$ on the imaginary axis at $T_c$ for various $1/u$.

Figs. 3(a,b) show the $n_i$ dependence of $T_c$ at $\delta = 0.25, 0.18$ for the values of $1/u$ assigned to Zn and Ni. The shape of the curves is essentially identical to the AG form (see for example [20]). The $T_c$ suppression rate is about two times as fast for $1/u = -0.1$ (Zn) than for $1/u = 0.7$ (Ni). Furthermore, it strongly increases for lower $\delta$, in qualitative agreement with experiments on $YBa_2Cu_3O_{7-y}$ [21]. The critical concentrations at $1/u = -0.1$ ($n_i^c \approx 0.06$ for $\delta = 0.25$, $n_i^c \approx 0.04$ for $\delta = 0.18$) also agree well with experimental values when taking into account the planar site preference of Zn, as discussed above. Fig. 3(c) illustrates the substantial increase in $dt_c/dn_i$ [$t_c = T_c/T_{c0}$, $T_{c0} = T_c(n_i = 0)$] upon lowering $\delta$. In analogy to $N_{res}$ [Fig. 1(a)], the peak shifts from negative to positive $1/u$ as $\delta$ is lowered. Fig. 3(d) shows the frequency dependence of the self-energy $\Sigma_0$ on the imaginary axis for various values of $1/u$ [Re$\{\Sigma_0(T_c)\} = 0$, and Im$\{\Sigma_3(T_c)\} = 0$]. The large variation of $dt_c/dn_i$ between $1/u = -0.1, 0.7$ originates in the big differences of $\Sigma_0$ particularly at intermediate $\omega_n$.

Finally, we need to check whether the $u$ values assigned to Zn and Ni can produce a quasiparticle damping Im$\Sigma_n$, which could explain the similarity in the residual resistivity caused by the two. The normal state dc conductivity is given by the Kubo formula (vertex corrections vanish for a contact impurity potential [22])

$$\sigma_{\text{imp}}^{xx} = \frac{\pi e^2}{\Lambda} \int d\omega \left(-\frac{\partial f}{\partial \omega}\right) \sum_{\mathbf{k}} \left[\frac{\partial \xi_{\mathbf{k}}}{\partial k_x} A(\mathbf{k},\omega)\right]^2, \quad (5)$$

where the spectral function is defined as

$$A(\mathbf{k},\omega) = -\frac{\text{sgn}\omega}{\pi} \frac{\text{Im}\Sigma_n}{(\omega - \xi_{\mathbf{k}} - \text{Re}\Sigma_n)^2 + (\text{Im}\Sigma_n)^2}, \quad (6)$$

and $f(\omega)$ the Fermi function. At low $T$, $\partial f/\partial \omega$ converges to a delta function, and only the self-energy around $\varepsilon_F$ is important. We assume that the 3D coupling in the cuprates is sufficient, so that localization effects can be neglected in the low density limit under consideration.

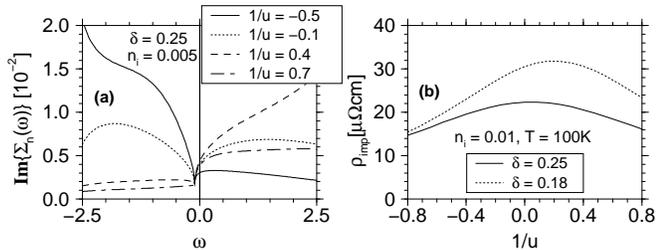

**Fig. 4.** (a) Frequency dependence of the normal state self-energy $\text{Im}\Sigma_n$ for various values of $1/u$. (b) Impurity induced resistivity at $T = 100K$, as a function of $1/u$ for $\delta = 0.18, 0.25$.

Fig. 4(a) shows the $\omega$ dependence of $\text{Im}\Sigma_n$ for $\delta = 0.25, n_i = 0.005$, and various values of $u$. The surprising observation is that even though the overall shape of the curves is different, in the vicinity of $\varepsilon_F$, they are very similar. In particular for $1/u = -0.1, 0.7$ the values at $\varepsilon_F$ are essentially the same. The vHs creates some type of an inversion point for the frequency dependent scattering rate. This similarity reflects itself in the impurity induced dc resistivity which we plot as a function of $1/u$, for $n_i = 0.01, T \approx 100K$, and $\delta = 0.25, 0.18$ in Fig. 4(b). The values between $1/u = -0.5, 0.7$ vary by less than 15%, even though the $T_c$ suppression within this range changes by a factor of two [see Fig. 3(c)]. The magnitude also agrees well with reported values for Zn and Ni ($\approx 20 - 30\mu\Omega$cm/at%) [6]. This shows that the proximity to the vHs, causing the strong $\omega$ dependence of $\Sigma_n$ is capable to *severely violate* the conventional proportionality of residual resistivity and $T_c$ suppression found for a constant DOS. The large resistivity caused by Ni together with its mild effect on $T_c$ as compared to Zn, might be explainable by this effect.

In conclusion, we have shown that a simple BCS model with a $d_{x^2-y^2}$ OP, supplemented by the experimental quasiparticle dispersion, and non-magnetic impurity scattering provides a consistent description of the experimentally observed effects of Zn and Ni. The presence of the vHs slightly below $\varepsilon_F$ is necessary to resolve the puzzle given by the violation of the proportionality between residual resistivity and $T_c$ suppression for the two types of impurities.

The author would like to thank M. R. Norman, B. Farid and R. Zeyher for useful comments, and acknowledges financial support by the NSF (DMR-91-20000) through the Science and Technology Center for Superconductivity. Part of this work was done at the Materials Science Division, Argonne National Laboratory, Argonne, Illinois 60439.